\documentclass[aps,prl,twocolumn,longbibliography]{revtex4}

\usepackage[utf8]{inputenc}
\usepackage{bm}
\usepackage{bbold}
\usepackage{hyperref}
\usepackage[normalem]{ulem}
\usepackage{color}
\usepackage[dvipsnames]{xcolor}
\usepackage{graphicx}
\usepackage{mathtools}

\begin{document}

\title{Multipartite entangled states in dipolar quantum simulators}
\author{Tommaso Comparin$^1$, Fabio Mezzacapo$^1$, Tommaso Roscilde$^1$}
\affiliation{$^1$ Univ Lyon, Ens de Lyon, CNRS, Laboratoire de Physique, F-69342 Lyon, France}

\begin{abstract}
The scalable production of multipartite entangled states in ensembles of qubits is a fundamental function of quantum devices, as such states are an essential resource both for fundamental studies on entanglement, as well as for applied tasks. Here we focus on the $U(1)$ symmetric Hamiltonians for qubits with dipolar interactions -- a model realized in several state-of-the-art quantum simulation platforms for lattice spin models, including Rydberg-atom arrays with resonant interactions. Making use of exact and variational simulations, we theoretically show that the non-equilibrium dynamics generated by this lattice spin Hamiltonian shares fundamental features with that of the one-axis-twisting model, namely the simplest interacting collective-spin model with $U(1)$ symmetry. The evolution governed by the dipolar Hamiltonian generates a cascade of multipartite entangled states --  spin-squeezed states, Schr\"odinger's cat states, and multi-component superpositions of coherent spin states. Investigating systems with up to $N=144$ qubits, we observe full scalability of the entanglement features of these states directly related to metrology, namely scalable spin squeezing at an evolution time ${\cal O}(N^{1/3})$; and Heisenberg scaling of sensitivity of the spin parity to global rotations for cat states reached at times ${\cal O}(N)$. 
Our results suggest that the native Hamiltonian dynamics of state-of-the-art quantum simulation platforms, such as Rydberg-atom arrays, can act as a robust source of multipartite entanglement.
\end{abstract}

\maketitle


\emph{Introduction.} Quantum entanglement \cite{Horodeckietal2009} is the distinctive feature of many-body quantum mechanics, at the root of its fundamental complexity as well as of its potential as a technological resource \cite{Preskill2012Arxiv,TothA2014,Pezze2018RMP}. Generic pure states in the Hilbert space have a large bipartite entanglement, captured by entanglement entropies of a subsystem that scale like the subsystem volume \cite{Page1993}; a similar scaling is observed in states which are obtained \emph{e.g.} by evolving initially non-entangled states with a generic interacting many-body Hamiltonian for a macroscopic time, leading to quantum thermalization \cite{Kaufmanetal2016}. Nonetheless a more specialized form of entanglement is widely recognized as a resource, namely \emph{certifiable multipartite} entanglement, in which 1) the number of inseparable degrees of freedom (a.k.a. entanglement depth \cite{SorensenM2001}) is as big as a macroscopic fraction of the system, and 2) such a depth can be efficiently estimated with criteria based on the measurement of a few observables. States of this kind allow for an efficient (\emph{i.e.} scalable) entanglement certification \cite{Guehne_2009}; and they represent the basis of quantum technology tasks, such as entanglement-assisted metrology \cite{TothA2014,Pezze2018RMP}. Therefore identifying robust protocols that lead to an \emph{efficient} and \emph{scalable} production of certifiable multipartite states -- namely of states with an entanglement depth scaling with the number $N$ of degrees of freedom, and in a time scaling polynomially with $N$ -- is a central task of modern quantum science and technology. In this work we show that scalable production of multipartite entangled states can be achieved in qubit ensembles with $U(1)$ symmetric \emph{dipolar} interactions, which are most prominently realized by Rydberg atoms with resonant interactions \cite{BrowaeysL2020} among other platforms \cite{Mosesetal2017,Kucskoetal2018,Chomazetal2022}. 
Making use of state-of-the-art time-dependent variational approaches, pushed to macroscopic evolution times, we show that two-dimensional lattices of qubits interacting with dipolar couplings for two spin components, initialized in a coherent spin state along an interaction axis, evade generic thermalization; and they develop paradigmatic examples of multipartite entangled states, namely spin-squeezed states \cite{Kitagawa1993PRA,Wineland1994PRA} and Schr\"odinger's cat states \cite{Agarwal1997PRA, Molmer1999PRL}. The dynamics of dipolar systems is found to exhibit a deep similarity to that of the paradigmatic model of collective-spin interactions, namely the one-axis-twisting (OAT) Hamiltonian \cite{Kitagawa1993PRA}. In particular we observe cat-like states in dipolar lattices for up to $N=144$ qubits -- {a remarkable observation in a system with non-mean-field interactions}. Our work paves the way for the scalable production of multipartite entanglement in dipolar quantum simulators. 

 \begin{figure*}[hbt]
\centering
\includegraphics[width=0.8\textwidth]{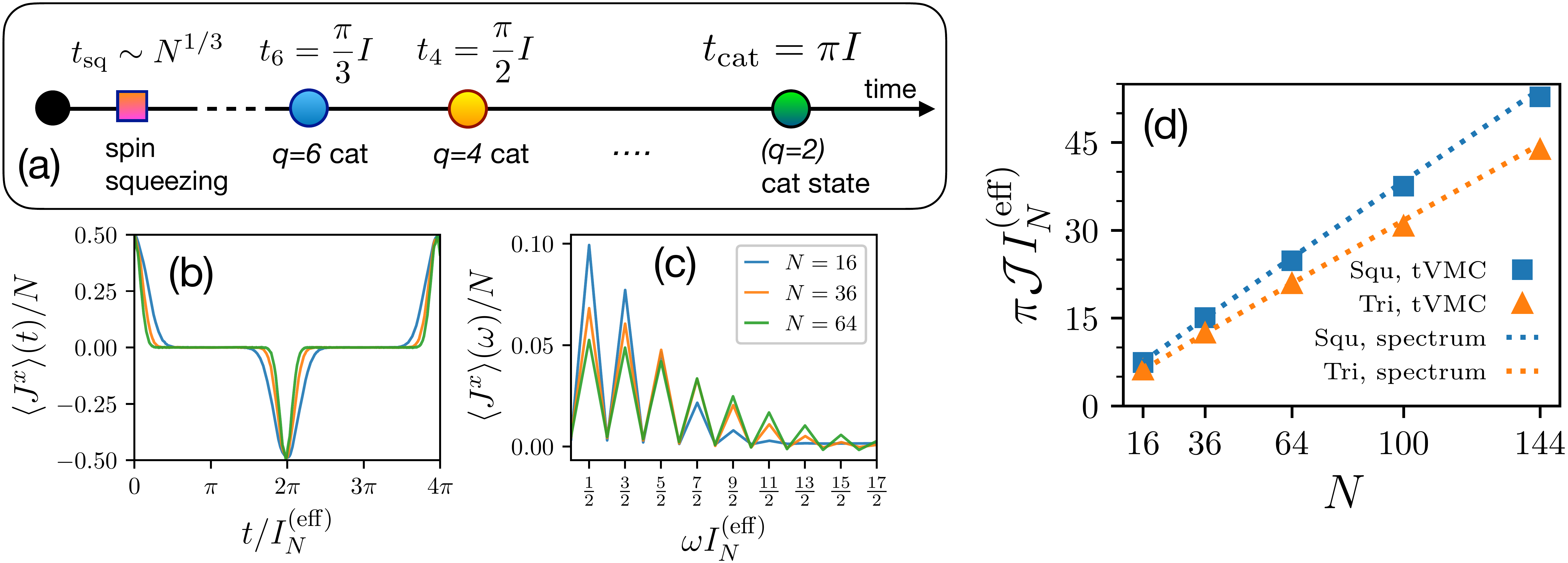}
\caption{(a) Cascade of entangled states observed in the OAT model and in this work -- here we only indicate even-headed cat states, but $q$-headed states with odd $q$ exist as well, at times $t = 2\pi I/q$; (b) Dynamics of the average spin $\langle J^x \rangle(t)$ for the dipolar XX model on the square lattice; (d) Fourier transform $\langle J^x(\omega) \rangle$; (d) effective moment of inertia for the square (Squ) and triangular (Tri) dipolar lattices as extracted from the tVMC dynamics (squares and triangles); and estimated from the low-energy spectrum (dashed lines). The specific combination $\pi I^{\rm (eff)}_N {\cal J}$ represents the time of the $q=2$ cat state.}
\label{fig:Jx}
\end{figure*}

\emph{Certifiable multipartite entanglement.} We specialize our attention to the case of qubit ensembles, whose most basic description is achieved in terms of the collective spin operator $\bm J = \sum_{i=1}^N {\bm S}_i$, where $\bm S_i$ represent spin-1/2 operators for each of the $N$ qubits. 
A primary example of certifiable multipartite entanglement in qubit ensembles is offered by spin-squeezed states \cite{Kitagawa1993PRA,Wineland1994PRA}, namely entangled states that are characterized by a finite net spin orientation $\langle \bm J \rangle$, and by relative fluctuations of one collective-spin component transverse to the average orientation which are reduced with respect to all fully separable states. The relative fluctuations are captured by the spin-squeezing parameter \cite{Wineland1994PRA} 
\begin{equation}
\xi_R^2 = \frac{N \min_\perp {\rm Var}(J^\perp)}{|\langle {\bm J}\rangle|^2}
\end{equation}
where the minimization is made over the plane perpendicular to $\langle \bm J \rangle$; and spin squeezing amounts to the condition $\xi_R^2 < 1$. 
Spin squeezing is an entanglement witness \cite{Sorensen2001} and it is an entanglement-depth estimator: $\xi_R^2<1/k$ with $k>1$ guarantees that a state is not $k$-producible, namely the smallest block of entangled qubits is composed of $k+1$ elements \cite{Pezze2009PRL,Hyllus2012PRA,Toth2012PRA}. Moreover, when subject to rotations $e^{-i\theta J'_{\perp}}$ around the anti-squeezed direction $J'_{\perp}$ (perpendicular to both  $\langle {\bm J}\rangle$ and to the squeezed component), spin squeezed states allow for an estimate of the angle of rotation with an uncertainty $\delta \theta = \xi_R/\sqrt{N}$ below the so-called standard quantum limit $(\delta \theta)_{\rm SQL} = 1/\sqrt{N}$.  A second example of certifiable multipartite entanglement is offered by Schr\"odinger's cat (or Greenberger-Horne-Zeilinger~-- GHZ \cite{Greenberger1989}) states. Introducing the coherent spin state (CSS) with all spins polarized along the ${\bm n}$ direction, $|{\rm CSS}_{\bm n}\rangle = | {\bm n}  \rangle^{\otimes^N}$  -- with $|\pm {\bm n} \rangle$ a generic qubit state with Bloch vector $\pm {\bm n}$ --  the most general form for a cat state (up to local unitaries) is   $|{\rm GHZ}_{\bm n} \rangle = \left ( | {\rm CSS}_{\bm n}\rangle + e^{i\phi} | {\rm CSS}_{-\bm n}\rangle \right ) /\sqrt{2}$. This state has an entanglement depth of $N$, and, when rotated around the ${\bm n}$ direction with the unitary $e^{-i\theta \bm J \cdot \bm n}$, it allows for an estimate of the rotation angle with uncertainty $\delta \theta = 1/N$, representing the ultimate (Heisenberg) limit for phase estimation. A generalization of the cat state is offered by so-called ``$q$-headed" cat states \cite{Agarwal1997PRA}, which are superpositions of $q$ CSS along directions ${\bm n}_p$ ($p = 0, ..., q-1$) in \emph{e.g.} the $xy$ plane, forming an angle of $2\pi p/q$ with the $x$ axis : $ |q{\rm-cat}\rangle  = {\cal A}^{-1} \sum_{p=0}^{q-1} c_p | {\rm CSS}_{{\bm n}_p} \rangle$ with complex $c_p$ coefficients of unit modulus, {and $\cal A$ a normalization factor.}

\emph{Long-range interacting XX Hamiltonians and OAT model.} In this work we show how spin-squeezed states and cat-like states are generated along the unitary dynamics initialized in the coherent spin state $|{\rm CSS}_x\rangle$ with ${\bm n} = \bm e_x$, and governed by the long-range XX ferromagnetic Hamiltonian
\begin{equation}
{\mathcal H}_{\alpha-{\rm XX}} = - \frac{\cal J}{{\cal N}_\alpha}
\sum_{i<j} \frac{1}{r_{ij}^\alpha} \left ( S_i^x S_j^x + S_i^y S_j^y \right )
\label{eq:H}
\end{equation}
where ${\mathcal J}>0$ is the coupling constant; $r_{ij}$ is the distance between the $i$-th and $j$-th spins; and ${\cal N}_\alpha$ is a normalization factor ensuring an extensive energy. Throughout this work we shall consider spins arranged on a planar lattice with $N=L\times L$ sites and periodic boundary conditions; we shall present results for both the square and the triangular lattice. All of our results are for dipolar ($\alpha=3$) interactions (for which we can take ${\cal N}_3 = 1$), so as to realize with Eq.~\eqref{eq:H} the Hamiltonian of resonantly interacting Rydberg atoms \cite{BrowaeysL2020}. The latter system has in fact antiferromagnetic interactions (${\mathcal J}<0$), but our focus is on the unitary dynamics initialized from a time-reversal symmetry state such as $|{\rm CSS}_{x}\rangle$, for which the global sign of the interactions is irrelevant \cite{Frerot2018PRL}. In order to understand the dynamics of the dipolar system, a fundamental reference is offered by the limit $\alpha=0$. In this limit, taking ${\cal N}_0 = N$, one obtains ${\mathcal H}_{0-{\rm XX}} = \frac{(J^z)^2}{2I} + c $, with $c = - ({\cal J}/{2N}) {\bm J}^2 + {\cal J}/4$ a constant factor, since the Hamiltonian commutes with ${\bm J}^2$. The latter Hamiltonian is the OAT model \cite{Kitagawa1993PRA} of a planar rotor with moment of inertia $I = N/{\cal J}$, whose dynamics is exactly solvable. When the dynamics is initialized in the $|{\rm CSS}_x\rangle$ state, this model is known to generate a cascade of entangled states  \cite{Kitagawa1993PRA, Agarwal1997PRA, Molmer1999PRL, Pezze2018RMP} (see Fig.~\ref{fig:Jx}(a) for a sketch), namely: 1) at a time $t_{\rm sq} \sim N^{1/3}$, a spin-squeezed states with $\xi_R^2 \sim N^{-2/3}$; 2) (with $N$ even) at times  $t_q = 2\pi I/q$ a $q$-cat state -- in particular a GHZ state of the kind $|{\rm GHZ}_x\rangle =(|{\rm CSS}_x\rangle + i |{\rm CSS}_{-x}\rangle)/\sqrt{2}$ for $t_{\rm GHZ} = \pi I$.  
The OAT Hamiltonian can be realized with spinor Bose condensates in a single spatial mode, and spin squeezing has been observed in seminal experiments \cite{Esteve2008, Riedel2010, Hosten2016} (see also \cite{Bohnet2016} for a trapped-ion realization); more recently its implementation with superconducting circuits has allowed the generation of ($q$-headed) cat states with up to $N=20$ qubits \cite{Song2019}. The full OAT dynamics is also realized with giant single-atom spins in Dy gases \cite{Chalopin2018NC}. The main result of this work is that the same sequence of entangled states can be realized with the dipolar Hamiltonian Eq.~\eqref{eq:H} with $\alpha=3$ for Rydberg atoms, with metrological qualities of the produced states that have the same scaling behavior as in the ideal case of the OAT dynamics. This result is far from trivial, as the OAT model is integrable (with non-thermalizing dynamics), while the dipolar Hamiltonian is expected to be chaotic (see discussion below).  

\emph{Time-dependent variational dynamics.} 
To investigate the scalable production of entangled states along the dynamics generated by the dipolar XX model, we compute the exact dynamics up to $N=20$ qubits~\cite{Weinberg2017SP, Weinberg2019SP}, and for larger $N$ we employ a time-dependent Variational Monte Carlo (tVMC) scheme~\cite{Carleo2012SR, Becca2017}, based on the pair-product (or spin-Jastrow) wavefunction ~\cite{Thibaut2019PRB} $ |\Psi(t)\rangle =: \sum_{\bm \sigma} \prod_{j\neq k} c_{jk}(\sigma_j, \sigma_k; t) |\bm \sigma\rangle$, where $\sigma_i$ is the state of the $i$-th spin on the computational basis (eigenbasis of $S_i^z$). 
The evolution of the pair coefficients $c_{jk}$ is dictated by the time-dependent variational principle. This wavefunction captures \emph{exactly} the dynamics of the OAT model \cite{Comparin2022PRA}; as shown in the Supplemental Material (SM) \cite{SM}, it remains extremely accurate in the case $\alpha=3$ on planar lattices, when compared with exact calculations for small lattices; and it allows us to push the calculation of the dynamics to sizes $N \sim 100$ and to reach macroscopic evolution times $t {\cal J} \sim {\cal O}(N)$ thanks to its small number of variational parameters (${\cal O}(N)$ with translational symmetry).

\emph{OAT-like dynamics of a planar dipolar array.} To establish a first link between the OAT dynamics and the dynamics of the dipolar XX model, we investigate the time evolution of the average collective spin, whose only component which is not identically zero is $\langle J^x \rangle(t)$. Fig.~\ref{fig:Jx}(b) shows the time evolution of $\langle J^x \rangle$, exhibiting the characteristic pattern of the OAT dynamics, with an inversion of the collective spin orientation at time $t_{\rm inv}$ followed by a revival of the original orientation at time $t_{\rm rev}$.  These two events occur at times $2\pi I$ and $4\pi I$ in the OAT dynamics, and therefore they allow us to define an \emph{effective}, size-dependent moment of inertia $I^{\rm (eff)}_N$ for the dipolar system such that $t_{\rm inv} = 2\pi I^{\rm (eff)}_N$ and $t_{\rm rev} = 4\pi I^{\rm (eff)}_N$.
The effective moment of inertia $I^{\rm (eff)}_N$ for the dipolar square and triangular lattices is shown in Fig.~\ref{fig:Jx}(c), and it is found to scale linearly with $N$; in particular the triangular lattice has a smaller $I^{\rm (eff)}_N$ due to its higher connectivity, guaranteeing a faster dynamics. 
In fact, as further discussed in the SM \cite{SM}, $I^{\rm (eff)}_N$ can be predicted \emph{ab-initio} by inspecting the low-energy excitation spectrum for a small system ($N=16$); and recognizing in it the characteristic planar rotor spectrum (known as Anderson tower of states \cite{Anderson1997,Tasaki2018JSP,Comparin2022PRA}). This allows us to extract the moment of inertia $I^{\rm (eff)}_{N=16}$, which can then be appropriately rescaled to an arbitrary size $N$ by using Kac renormalization factors, in very good agreement with the moment of inertia extracted directly from the time dependence of system of size $N$ (see Fig.~\ref{fig:Jx}(c)).
The Fourier transform of $\langle J^x \rangle(t)$ further reveals the nature of the low-lying energy spectrum of the system as that of a planar rotor: indeed, as $J^x$ connects states with $J^z=M$ differing by one unit, one expects \cite{SM} to see characteristic frequencies with energies $\omega I_N^{\rm (eff)}= [(M+1)^2-M^2]/2 = M + 1/2$, which is precisely what is observed in Fig.~\ref{fig:Jx}(c).  

\begin{figure}[hbt]
\centering
\includegraphics[width=0.98\linewidth]{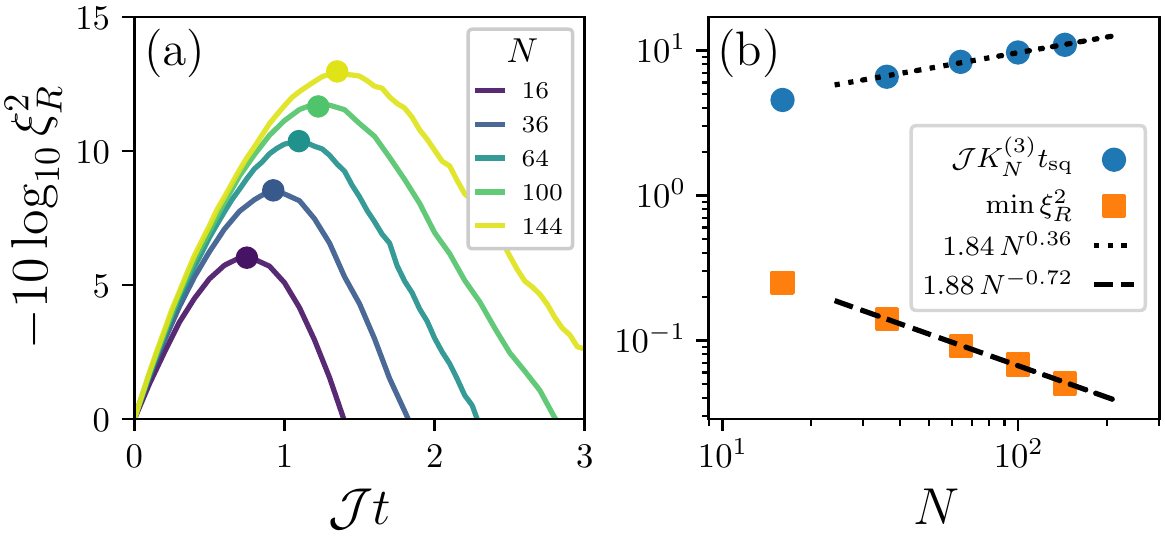}
\caption{(a) Evolution of the spin squeezing parameter for the dipolar XX model on a square lattice -- the circles mark the optimum; (b) scaling of the optimal squeezing value and optimal squeezing time (with Kac normalization $K_N^{(\alpha)}$ \cite{SM}), showing exponents $\nu=0.72$ and $\mu=0.36$ (to be compared with $\nu=2/3$ and $\mu=1/3$ for the OAT model).}
\label{fig:squeezing}
\end{figure}

\emph{Squeezed states and OAT scaling.} The first class of multipartite entangled states produced by the Hamiltonian dynamics is represented by spin squeezed states: Fig.~\ref{fig:squeezing} shows the time evolution for the squeezing parameter for various system sizes: clearly scalable squeezing is exhibited, with optimal squeezing time and optimal squeezing {scaling in a way which is compatible with the behavior of OAT model}. {Our results are consistent with those of Ref.~\cite{Perlin2020PRL}, based on an independent semiclassical calculation.} 

\begin{figure}[htb]
\centering
\includegraphics[width=\linewidth]{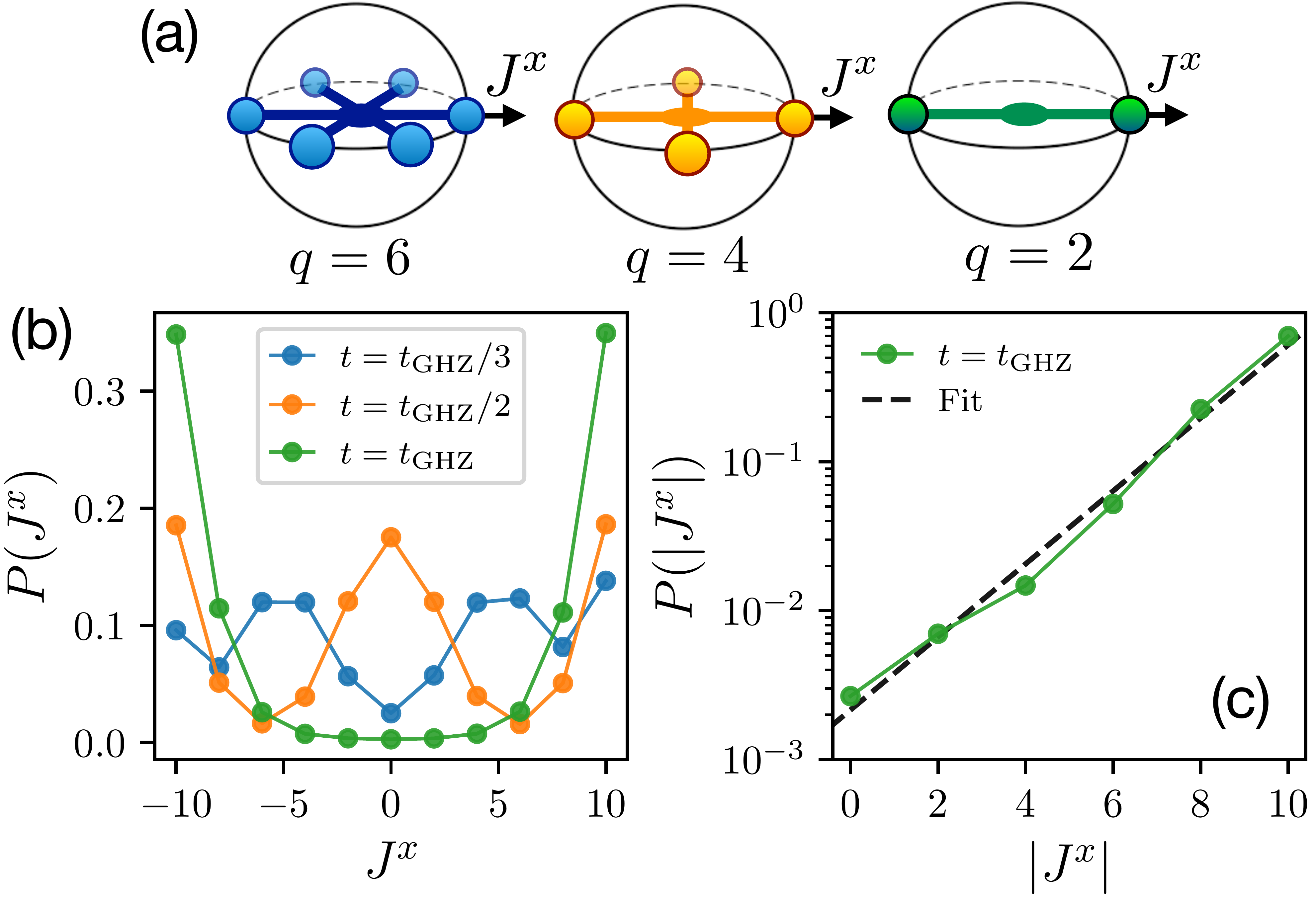}
\caption{(a) Sketch of the $q$-cat states studied here -- the balls indicate the coherent states involved in the $q$-cat state; (b) Distributions $P(J^x)$ obtained at the times of formation of the cat states with $q=6$, 4 and 2, obtained via exact calculations on a $5\times 4$ square lattice  -- only the $P$ values for even $J^x$ are indicated, as the ones for odd $J^x$ vanish identically, because the Hamiltonian is parity conserving. (c) Log-lin plot of the tail of the peaks for the $q=2$ cat state; the dashed line is the fit to an exponential.}
\label{fig:FCS_Jx}
\end{figure}

\begin{figure}[hbt]
\centering
\includegraphics[width=\linewidth]{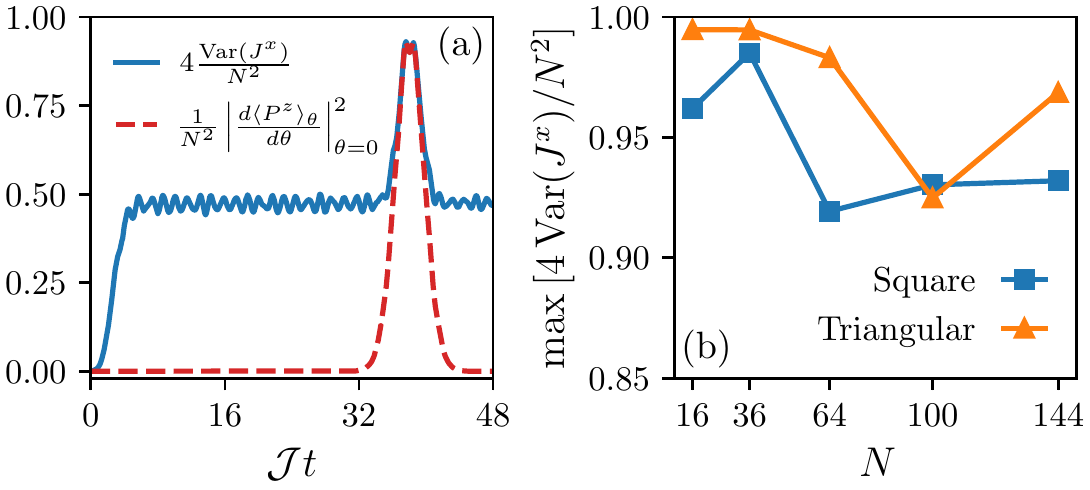}
\caption{(a) Time evolution of $4{\rm Var}(J^x)/N^2$ compared to that of the inverse uncertainty on phase estimation from the evolution of the parity -- the results are for a square lattice with $N=100$; (b) Size dependence of the maximal variance of $J^x$ during time evolution, for both square or triangular lattices.}
\label{fig:info}
\end{figure}

\emph{Multi-headed and double-headed cat states.} The squeezing dynamics is followed by the generation of over-squeezed states: their entanglement pattern is best recognizable at times $2 \pi I^{\rm (eff)}/q$, at which these states are expected to take the form of $q$-cats (see Fig.~\ref{fig:FCS_Jx}(a) for a sketch). In order to detect the appearance of a $q$-cat, we inspect the probability distribution $P(J^x)$ for the $J^x$ spin component \cite{Ferrini2009PRA}, reconstructed via exact calculations in Fig.~\ref{fig:FCS_Jx}(b) {(while in \cite{SM} we show a tVMC study of the overlap with the $|q{\rm -cat}\rangle$ states)}. At times $2\pi I^{\rm(eff)}/q$ the $P(J^x)$ distribution exhibits a multi-peaked structure, reflecting the appearance of a $q$-cat as superposition of several CSS with discrete projections along the $J^x$ axis. In particular we observe a characteristic 4-peak structure for the $q=6$ cat state, a 3-peak structure for the $q=4$ cat state, and a 2-peak structure for the $q=2$ cat / GHZ state. In the latter case, the distribution associated with the ideal cat state would be $P(J^x)=1/2$ for $J^x=\pm N/2$ and zero otherwise, while the dipolar cat state exhibits instead two peaks with a tail. Nonetheless, as shown in Fig.~\ref{fig:FCS_Jx}(c), the tail in question decays exponentially when moving away from the maxima; this localized structure of the distribution around the maxima has important consequences that we shall further explore below. 

In spite of their different multi-peak structures, the distributions for the $q>2$ cat states have nearly the same variance, as shown in Fig.~\ref{fig:info}(a) -- therefore their specific nature is only seen via higher moments. On the other hand the $q=2$ cat/GHZ state stands out for its variance ${\rm Var}(J^x)$, which attains the maximum possible value of $N^2/4$ for $N$ qubits in the case of the ideal cat state, while it attains a value which approaches this maximum in the case of the state generated by the dipolar dynamics. As shown in Fig.~\ref{fig:info}(b), {for the system sizes of interest} the maximum of the variance ${\rm Var}(J^x)$ reached for dipolar cat states scales indeed with $N^2$, attaining a value which is $> 90\%$ of its maximum. Even though we do not have access to the $P(J^x)$ distribution within the tVMC approach, this result is fully coherent with the distribution remaining exponentially localized around $\pm N/2$ values up to the largest systems we considered.

\begin{figure}[hbt]
\centering
\includegraphics[width=0.95\columnwidth]{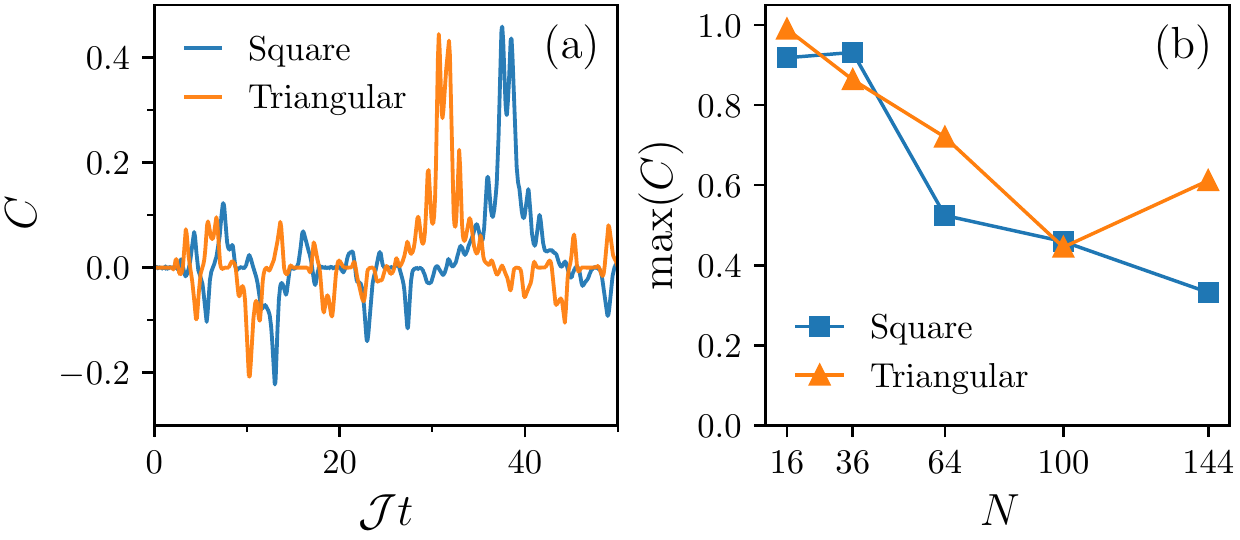}
\caption{(a) Many-body coherence $C =  2{\cal F}_{\rm GHZ} - {\cal F}_x - {\cal F}_{-x}$ as a function of time, for $N=100$ square and triangular lattices. 
(b) Size scaling of the maximal value of $C$ for both lattices.}
\label{fig:fidelity}
\end{figure}

\emph{Fidelity between the dipolar cat states and ideal ones}. A general figure of merit used for the realization of cat states in experiments (or in any imperfect preparation protocol) is their fidelity with respect to the desired target state \cite{Omran2019, Song2019}. Nonetheless this criterion should be used here with a grain of salt: indeed the preparation of an ideal cat state is \emph{not} the target of the evolution we study, given that it is not guaranteed even in the ideal conditions of the theoretical calculations. tVMC gives us access to the fidelity  ${\cal F}_{\rm GHZ} = |\langle {\rm GHZ}_x| \Psi(t)\rangle|^2$, as well to the fidelities ${\cal F}_{\pm x}$ with the two coherent spin states $|{\rm CSS}_{\pm x}\rangle$ \cite{SM}. In particular the difference $C = 2{\cal F}_{\rm GHZ} - {\cal F}_x - {\cal F}_{-x} = \langle |{\rm CSS}_{x}\rangle \langle {\rm CSS}_{-x}| + {\rm h.c.} \rangle$ is most relevant, as it probes the presence of $N$-body coherence in the system. Fig.~\ref{fig:fidelity}(a) shows that such coherence is maximal at the time of formation of the dipolar cat, but it is close to its maximum value of 1 (attained by an ideal cat state) only on small system sizes, while it is reduced to $\sim 0.5$ for the largest system sizes we considered. This observation is compatible with the result seen above that dipolar cat states are states with several $J^x$ components exponentially localized around the extremal $\pm N/2$ values; and therefore containing coherences between macroscopically distinct states other than $|{\rm CSS}_{\pm x}\rangle$. The fact that the state retains macroscopic quantum coherence, even though it deviates from the ideal GHZ state, is at the root of its extreme metrological sensitivity, as we shall see below. 

\emph{Heisenberg-limited interferometry using parity.} Similar to $\xi_R^2$ for squeezed states, a fundamental figure of merit for the entanglement content of cat-like states is provided by their sensitivity to rotations $U(\theta) = e^{-i\theta J^x}$, which is best captured by the $\theta$-dependence of the expectation value of the parity operator $P^z = \prod_i (2 S_i^z)$, namely of $\langle P^z \rangle_\theta = \langle \Psi(t) | U(-\theta) P^z U(\theta) | \Psi(t)  \rangle$. The quantum Cram\'er-Rao bound \cite{Pezze2018RMP} imposes that 
\begin{equation}
\max_{\theta*} \frac{1}{{\rm Var}(P^z)_{\theta=\theta^*}}
\left|
\frac{d\langle P^z\rangle_\theta}{d\theta}
\right|_{\theta=\theta^*}^2
\leq {\rm QFI}(J^x) \leq
4 {\rm Var}(J^x),
\label{eq:info}
\end{equation}
where the left-hand side expresses the inverse squared uncertainty $(\delta\theta)^{-2}$ on the angle estimation using the parity measurement, and ${\rm QFI}(J^x)$ is the quantum Fisher information associated with the $J^x$ operator, which in the case of pure states coincides with the upper bound given by $4 {\rm Var}(J^x)$. Our tVMC calculations allow us to reconstruct the left-hand side of the inequality Eq.~\eqref{eq:info} for $\theta^*=0$ \cite{SM}. The result is shown in Fig.~\ref{fig:info}(a) and compared to $4{\rm Var}(J^x)$: there we see that, upon formation of the dipolar cat state, the inequality chain of Eq.~\eqref{eq:info} collapses to an identity, showing that the measurement of the parity around $\theta^*=0$ is optimal -- as expected for cat states. This optimality is observed for all the system sizes we considered: therefore the fact that  $4 {\rm Var}(J^x) \approx a N^2$ with $a\gtrsim 0.9$ (as shown in Fig.~\ref{fig:info}(b)) allows us to conclude that dipolar cat states allow interferometry to attain the Heisenberg scaling; and to achieve $>90\%$ of the Heisenberg limit. 

\emph{Discussion and conclusions.} We have shown that planar qubit arrays with dipolar interactions can reproduce the entanglement dynamics of the one-axis twisting Hamiltonian, with the scalable production of spin squeezed states and of cat-like states. This result is rooted in the deep correspondence between the low-energy spectra of the two models, taking the form of a tower of states for a planar rotor. Nonetheless the cascade of entangled states and the revivals of the initial state observed in this work are in clear contradiction with the picture of quantum thermalization of closed quantum systems \cite{dalessio_quantum_2016}, in which local observables should exhibit small fluctuations around their thermodynamic equilibrium value; and this in spite of the fact that the dipolar spin model is expected to be a non-integrable one. The deviation of the observed dynamics with respect to standard thermalization can be understood within a picture in which the collective spin and the fluctuations of the spins at finite momentum effectively decouple, as we shall present in a forthcoming publication; nonetheless this decoupling is only approximate, and should break down at sufficiently long times. Yet our observation is that dipolar systems comprising $N \sim {\cal O}(100)$ qubits -- currently accessible experimentally with Rydberg-atom arrays \cite{Scholl2021,Ebadi2021} --  do not show any significant degradation of the decoupling picture up to macroscopic evolution times $\sim {\cal O}(N)$. This fundamental property of dipolar Hamiltonians implies that atomic quantum simulators realizing dipolar qubit ensembles with $U(1)$ symmetry -- Rydberg atoms \cite{BrowaeysL2020}, as well as dipolar molecules \cite{Mosesetal2017}, trapped ions \cite{Bohnet2016,Brydges2019}, magnetic atoms \cite{Chomazetal2022} etc. -- have the potential to reach unprecedented levels of multipartite entanglement, including cat states with $N>100$, and Heisenberg scaling of metrological properties. The latter scaling requires the measurement of the parity, which is perfectly accessible in state-of-the-art quantum simulators granting single-qubit addressability. 
In the specific case of Rydberg atoms with resonant interactions, the $\sim {\cal O}(N)$ evolution times required to reach large cat states may appear out of reach due to the finite lifetime of the Rydberg states; but the lifetime can be extended far beyond the requirements of our observations when using \emph{e.g.} circular Rydberg states \cite{Nguyenetal2018}. 

\begin{acknowledgments}
\emph{Acknowledgements.} We acknowledge useful discussions with A. Browaeys and I. Fr\'erot.
This work is supported by the Agence Nationale de la Recherche (EELS project, ANR-18-CE47-0004) and by QuantERA (``MAQS" project). All numerical simulations have been performed on the PSMN cluster of the ENS of Lyon. Exact results have been obtained through the QuSpin package~\cite{Weinberg2017SP, Weinberg2019SP}.
\end{acknowledgments}

\bibliography{refs.bib}


\clearpage

\appendix
\begin{center}
{\bf Supplemental Material} \\
{\bf \emph{Multipartite entangled states in dipolar quantum simulators}} 
\end{center}

\section{Benchmark of the pair-product Ansatz}

\begin{figure}[htb]
\centering
\includegraphics[width=0.98\linewidth]{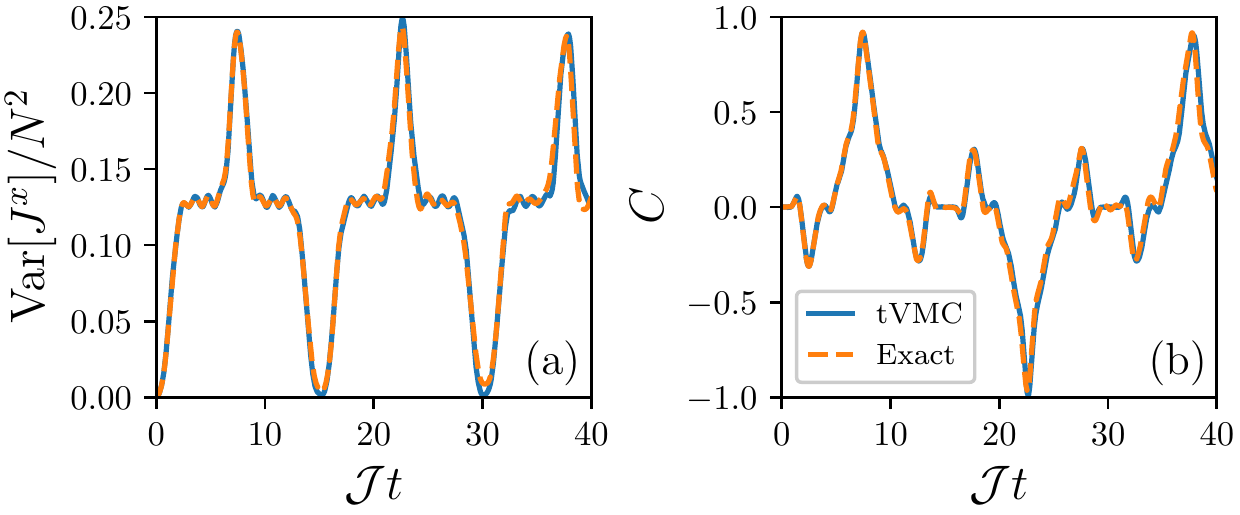}
\caption{Comparison between exact and tVMC dynamics for a $4\times4$ dipolar system on the square lattice. (a) ${\rm Var}(J^x)$; (b) many-body coherence $C$.}
\label{fig:benchmark}
\end{figure}

The main limitation of the tVMC method is its bias due to the specific choice of an Ansatz wavefunction. Nonetheless the pair-product Ansatz used in this work reproduces \emph{exactly} the dynamics of the one-axis-twisting (OAT) Hamiltonian (namely the long-range XX model with $\alpha=0$) starting from the state $|{\rm CSS}_x\rangle$; and therefore it is expected to remain highly accurate for sufficiently small values of $\alpha$ -- as already verified for one-dimensional systems~\cite{Comparin2022PRA}. Working in two dimensions (as in this work) makes the Ansatz even more accurate (for the same value of $\alpha$), because of the higher connectivity of the lattice which brings the system closer to the OAT case.  We put the quantitative accuracy of the variational dynamics to the test by comparing it with the exact dynamics on a small system (a periodic $4\times4$ square lattice), for which exact calculations can be performed.  Fig.~\ref{fig:benchmark} shows this comparison for the evolution of ${\rm Var}(J^x)$, and of the macroscopic coherence $C = 2{\cal F}_{\rm GHZ} - {\cal F}_x - {\cal F}_{-x}$: the variational time evolution is in excellent agreement with the exact one, with small deviations observed only at the large ($\sim{\cal O}(N)$) times. This result leads us to conclude that our variational results for larger systems sizes are fully quantitative.

\begin{figure}[htb]
\centering
\includegraphics[width=\linewidth]{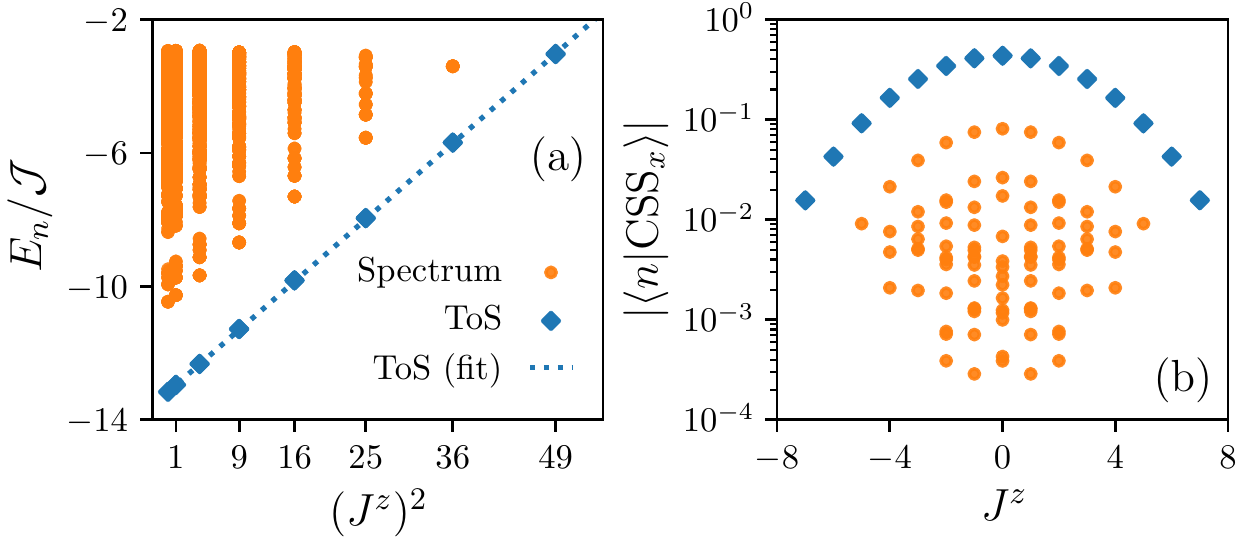}
\caption{(a) Low-energy spectrum of the dipolar XX model on a $4\times4$ periodic square lattice, resolved as a function of $J^z$. States of the ToS are marked with blue diamonds, and the dotted line is a linear fit of their energy as a function of $(J^z)^2$.
(b) Overlap of the eigenstates with the initial state of the dynamics (CSS$_x$ state).
}
\label{fig:tos}
\end{figure}

\section{Tower of states and effective OAT dynamics}

We consider the dipolar XX model on a periodic square lattice (with $N=4\times 4$ spins), and we compute the low-energy common eigenstates $|n\rangle$ of ${\cal H}$ and $J^z$.
As shown in Fig.~\ref{fig:tos}(a), an Anderson tower of states (ToS) is clearly identified in the spectrum as a set of states with energy scaling as $(J^z)^2$, and we extract an effective moment of inertia $I^{\rm (eff)}_{N=16}$ by fitting these energies to ${\rm const} + (J^z)^2 / (2 I^{\rm (eff)}_{N=16})$. This value reads
\begin{align}
{\cal J} I^{\rm (eff)}_{N=16} & = 2.4168 \qquad \text{(square lattice)},\\
{\cal J} I^{\rm (eff)}_{N=16} & = 1.9587 \qquad \text{(triangular lattice)}.
\end{align}
The value for the triangular lattice has been extracted in the same way by fitting the spectrum of a $4\times4$ periodic lattice. 

Knowing an estimate of $I^{\rm (eff)}_{N}$ for a given system size $N$ allows us to provide an estimate for larger system sizes $N'$, in spite of the fact that their diagonalization might be technically impossible. A scaling formula for the effective moment of inertia can be obtained by postulating that the dipolar model with $\alpha=3$ has the same behavior of a OAT Hamiltonian ($\alpha=0$) after proper normalization of the coupling constant. Such a normalization is provided by the $\alpha$-dependent Kac factor 
\begin{equation}
K_N^{(\alpha)} = \frac{1}{N} \sum_{i\neq j} \frac{1}{r_{ij}^\alpha} 
\end{equation}
which allow us to define a Kac-normalized long-range XX Hamiltonian
\begin{equation}
\tilde{\mathcal H}_{\alpha-{\rm XX}} = - \frac{\cal J}{K^{(\alpha)}_N}
\sum_{i<j} \frac{1}{r_{ij}^\alpha} \left ( S_i^x S_j^x + S_i^y S_j^y \right )~.
\label{eq:HKac}
\end{equation}
The moment of inertia extracted above for the dipolar model is related to that of the Kac-normalized version of the same model by the relationship $\tilde{I}^{\rm (eff)}_N = {I}^{\rm (eff)}_N K_N^{(3)}$.

In the case of the OAT limit $\alpha=0$, one can immediately observe that the (Kac-normalized) moment of inertia reads $\tilde{I}_N = K_N^{(0)}/{\cal J}$, so that one obtains the scaling relation
\begin{equation}
\tilde{I}_N = \frac{K_{N}^{(0)}}{K_{N'}^{(0)}} \tilde{I}_{N'}~.
\end{equation}
We postulate then that the \emph{effective} moment of inertia of the Kac-normalized dipolar model scales in the same way as for the Kac-normalized OAT model, namely
\begin{equation}
\tilde{I}_{N}^{\rm (eff)} =  \frac{K_{N}^{(0)}}{K_{N'}^{(0)}}  \tilde{I}_N^{\rm (eff)}  \Rightarrow I_{N}^{\rm (eff)} = \frac{K_{N'}^{(3)}}{K_{N}^{(3)}} \frac{K_{N}^{(0)}}{K_{N'}^{(0)}} I_{N'}^{\rm (eff)}~.
\end{equation}
Using the reference value $N'=16$, for which the moment of inertia could be extracted from exact diagonalization, we can then reconstruct the effective moment of inertia from the above scaling formula for all the other system sizes we studied up to $N=144$. The results of the above scaling formula are shown in Fig.~1 of the main text, and they are in very good agreement with the estimate of the effective moment of inertia extracted from the tVMC dynamics (specifically from the time of the dipolar cat, estimated as $t_{\rm GHZ} = \pi I_N^{\rm (eff)}$). 

\begin{figure*}[htb]
\centering
\includegraphics[width=0.98\linewidth]{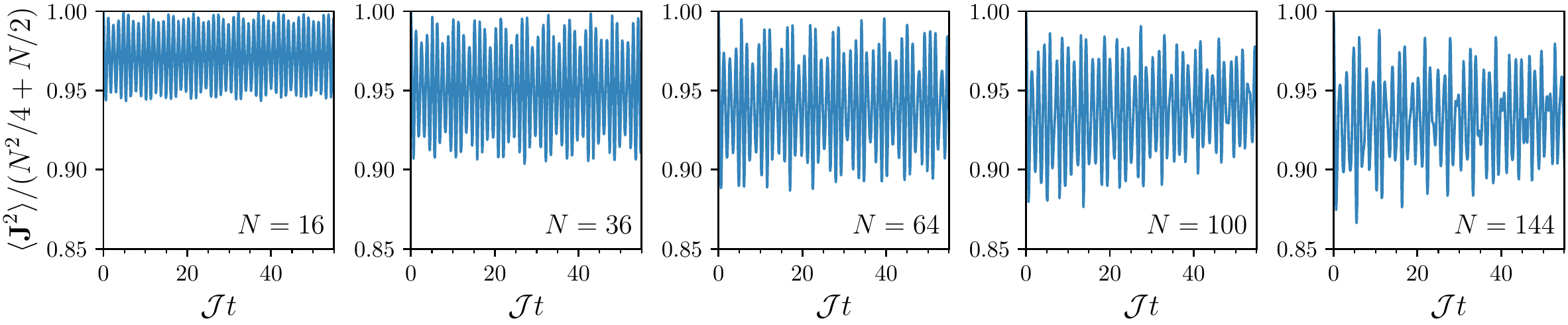}
\caption{Dynamics of the total spin $\langle {\bm J}^2\rangle$ for the dipolar XX model on the periodic square lattice, obtained by tVMC. Different panels correspond to different system sizes.}
\label{fig:totalspin}
\end{figure*}

\begin{figure}[htb]
\centering
\includegraphics[width=0.75\linewidth]{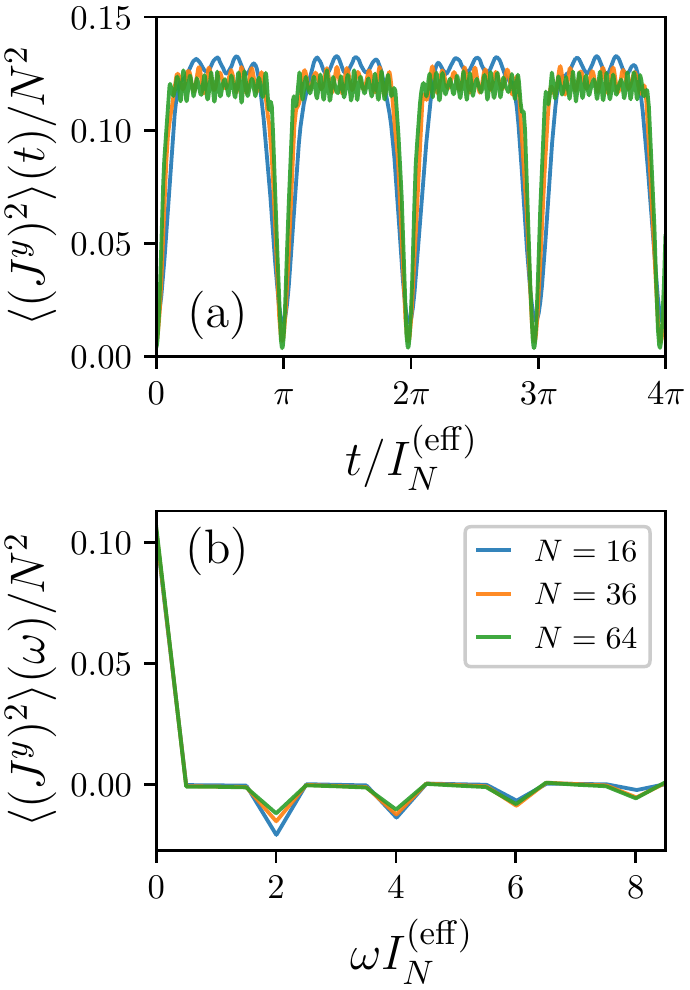}
\caption{Collective-spin dynamics in (a) real time and (b) frequency space for $\langle (J^y)^2 \rangle$, obtained via tVMC for the dipolar XX model on a square periodic lattice. Different lines correspond to different system sizes.}
\label{fig:tos_spectroscopy}
\end{figure}

\section{Dynamics of the spin length, and quench spectroscopy of the tower of states}

The crucial role of the tower of states in the dipolar XX dynamics follows from the fact that the time-evolved state remains close to the sector of maximal total spin. The initial state $|{\rm CSS}_x\rangle$ clearly has the strongest overlaps with the ToS states -- see Fig.~\ref{fig:tos}(b). At later times, the total squared spin deviates only slightly from its initial (maximum) value, $\langle {\bm J}^2\rangle(t=0)=(N/2)(N/2+1)$, retaining $\gtrsim 90\%$ of its length without any appreciable scaling for $N \geq 64$ -- see Fig.~\ref{fig:totalspin}.

The above result shows that the dynamics remains approximately confined to the sector with maximal total spin length, composed by the so-called Dicke states 
$|J=N/2,M\rangle$ with $M = -N/2, ... , N/2$. This observation is at the root of the similarity between the dipolar dynamics and the OAT dynamics (which remains strictly confined in the Dicke-state sector), and it has other significant consequences on the dynamics of the collective spin operator ${\bm J}$. 
In general, time-evolution of the expectation value of an operator $O$ can be written as
\begin{equation}
\langle O \rangle (t) = \sum_{nm} e^{i(E_m-E_n) t} \langle m | O | n\rangle c_n c^*_m
\end{equation}
where $|n\rangle$ , $|m\rangle$ are Hamiltonian eigenstates, with related eigenvalues $E_n, E_m$, and $c_n = \langle n| \Psi(0)\rangle$ is the overlap of the initial state with the related Hamiltonian eigenvector. The Fourier transform $\langle O \rangle (\omega) = \int dt e^{i\omega t} \langle O \rangle (t) $ reads therefore
\begin{equation}
 \langle O \rangle (\omega) = 2\pi \sum_{nm}   c_n c^*_m \langle m | O | n\rangle  \delta (\omega - \omega_{nm})~
\end{equation}
where $\omega_{nm} = E_n - E_m$, and we have set $\hbar =1$. 
If $O = J^x$ and for a dynamics confined to Dicke states, we have that 
\begin{align}
\langle J, M | J^x | J, M'\rangle = \frac{1}{2} &\Big(   \sqrt{J(J+1) - M(M-1)} ~\delta_{M',M-1} \nonumber \\
 &+  \sqrt{J(J+1) - M(M+1)} ~\delta_{M',M+1} ~\Big ) 
\end{align}
so that only transitions $M \to M\pm 1$ contribute to the spectrum of $\langle O \rangle(t)$. For a OAT model, or alternatively for states belonging to the ToS, one has that $\omega_{M\pm1, M} I= \pm (M + \frac{1}{2})$, with spectral weight decreasing with $|M|$ via the form factor $\langle m | O | n\rangle$ as well as via the overlaps $c^*_{M} c_{M\pm1}$. This behavior is clearly observed in Fig.~1(c) of the main text in the dipolar model, when trading $I$ for $I^{\rm (eff)}$ . If instead one uses $O= (J^y)^2 = \frac{1}{4} \left ( J^+ J^- + J^- J^+ - (J^+)^2 - (J^-)^2 \right )$, only transitions $M\to M$ or $M \to M\pm 2$ contribute to the Fourier spectrum (with a negative spectral weight), with frequencies $\omega_{M,M}=0$ or $\omega_{M\pm 2, M} I= \pm 2(M+1)$, respectively. This is indeed observed in Fig.~\ref{fig:tos_spectroscopy}, once again with the identification $I \to I^{\rm (eff)}$. 
These results show in turn that the Fourier analysis of the collective-spin dynamics allows one to spectroscopically reconstruct the ToS in the low-lying spectrum of a system.

\section{Computing fidelities in VMC}

For two (non-normalized) quantum states $|\Phi\rangle$ and $|\Psi\rangle$, their squared overlap can be written as
\begin{equation}
\frac
{|\langle \Phi | \Psi \rangle|^2}
{\langle \Phi | \Phi \rangle \langle \Psi | \Psi \rangle
} =
\left \langle
\frac{\Psi({\bm \sigma})}{\Phi({\bm \sigma})}
\right\rangle_{\Phi}
\left \langle
\frac{\Phi({\bm \sigma})}{\Psi({\bm \sigma})}
\right\rangle_{\Psi},
\end{equation}

where $\langle \dots\rangle_{\Phi}$ and $\langle \dots\rangle_{\Psi}$ are averages over the probability distributions $|\Phi({\bm \sigma})|^2/\langle \Phi | \Phi \rangle$ and $|\Psi({\bm \sigma})|^2/\langle \Psi | \Psi \rangle$.
When $|\Phi\rangle$ and $|\Psi\rangle$ can be represented via a variational Ansatz amenable to Monte Carlo sampling (like the pair-product wave function used in this study), the overlap can be obtained as the product of two Monte Carlo averages.  A similar estimator for fidelities was recently used in Ref.~\cite{Medvidovic2021NPJQI}.

In order to compute fidelities of the time-evolved state $|\Psi(t)\rangle$ with respect to some reference states, it is convenient to express all states in the same variational representation. In particular in our work we are interested in the fidelities ${\cal F}_{\rm GHZ}(t)$, ${\cal F}_x(t)$ and ${\cal F}_{-x}(t)$ of the evolved state with a GHZ state and with coherent spin states aligned along $x$ and $-x$. 
A general CSS with collective spin pointing in the $\theta$ direction in the \emph{xy} plane [that is, with $\langle (\cos\theta) J^x + (\sin\theta) J^y \rangle = N/2$] corresponds to the following (non-normalized) pair-product state,
\begin{equation}
\Psi_{\theta}({\bm \sigma}) =
\prod_{j<k}
\exp\left [ \frac{-i\theta (\sigma_j + \sigma_k)}{4(N-1)}  \right]~.
\end{equation}
The GHZ state $|{\rm CSS}_x\rangle + i |{\rm CSS}_{-x}\rangle$ can also be expressed in a (non-normalized) pair-product form,
\begin{equation}
\langle {\bm \sigma} | {\rm GHZ}_x \rangle =
\prod_{j<k}
\exp\left [ -i\frac{\pi}{2} \delta_{\sigma_j,\sigma_k} \right].
\label{e.ghz_x}
\end{equation}
This is the GHZ state that appears during OAT or dipolar dynamics when $N$ is an integer multiple of 4; the GHZ state appearing for $N=2,6,10,..$ (namely $|{\rm CSS}_x\rangle - i |{\rm CSS}_{-x}\rangle$) is obtained by changing the sign of the exponent in Eq.~\eqref{e.ghz_x}.
Note that any other state appearing during the OAT dynamics for a given $N$ can be directly written in a pair-product representation, since such an Ansatz is exact for this model; the explicit expression is derived in Ref.~\cite{Comparin2022PRA}. This allows us to compute fidelities with the finite-$N$ $q$-cat states -- see next section.

\section{Large-$N$ multicomponent GHZ states in dipolar XX dynamics}

\begin{figure}[htb]
\centering
\includegraphics[width=0.8\linewidth]{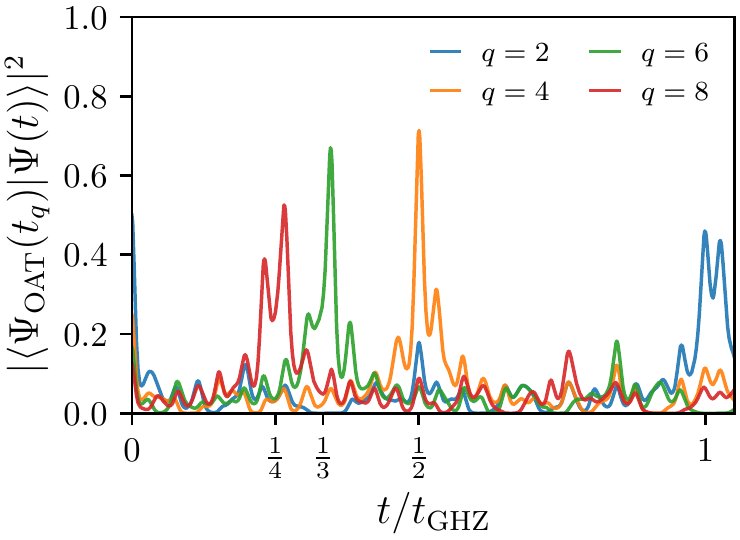}
\caption{Overlap between the states $| {\Psi}_{\rm OAT}(t_q)\rangle$ obtained via the OAT dynamics at times $t_q I = 2\pi/q$ and the state obtained along the dynamics governed by the dipolar Hamiltonian, for a square lattice with $N = 10\times 10$. Time is shown in units of $t_{\rm GHZ}$, which we take to be the time of maximal fidelity with the $q=2$ GHZ state.
Peaks are visible at times $t=t_{\rm GHZ}$ (for $q=2$), $t=t_{\rm GHZ}/2$ (for $q=4$), $t=t_{\rm GHZ}/3$ (for $q=6$), and $t=t_{\rm GHZ}/4$ (for $q=8$) . }
\label{fig:qcats_fidelities}
\end{figure}

As already mentioned in the main text, the dynamics of the OAT model generates special states at times $t_q = 2 t_{\rm GHZ}/q$, which are the linear combination of $q$ coherent spin states $|{\rm CSS}_{{\bm n}_p} \rangle$, forming angles $2\pi p/q$ with the $x$ axis, where $p=0,\dots,q-1$. The exact results on the distribution $P(J^x)$ in the main text (for $N=20$) show that such states also appear in the dynamics generated by the dipolar XX model.

Our tVMC calculations allow us to reach sizes exceeding the ones accessible with exact calculations. Yet we are unable to reconstruct the $P(J^x)$ distribution from tVMC, since the pair-product Ansatz is formulated in a basis in which $J^x$ is off-diagonal. Nonetheless we can still probe the formation of $q$-cat states on sizes far exceeding the realm of exact diagonalization, by computing the overlap of the time-evolved state $|\Psi(t)\rangle$ with the $q$-cat states appearing in the evolution of the OAT Hamiltonian for the same system size. Indeed the latter evolution is perfectly captured by the pair-product Ansatz, so that overlaps can be calculated as explained in the previous section. 

Fig.~\ref{fig:qcats_fidelities} shows the squared overlap $|\langle {\Psi}_{\rm OAT}(t_q) | \Psi(t)\rangle|^2$ between the $q$-cat state realized by the OAT dynamics at times $t_q$, $| {\Psi}_{\rm OAT}(t_q)\rangle$ (for which an exact representation as pair-product state exists~\cite{Comparin2022PRA}); and the tVMC result for the state $|\Psi(t)\rangle$ generated by the evolution governed by the dipolar Hamiltonian, the size $N$ being the same. For a system with $N = 100$, we observe a well ordered structure of peaks in the overlaps corresponding to times $t_q$ for the dipolar system having the same structure of those for the OAT model. This result shows therefore that the dipolar dynamics allows for a scalable production of $q$-cat states. For the $q$ CSS to be well resolvable in the $q$-cat state one needs that they form an angle $2\pi/q$ larger than the angular uncertainty on the orientation intrinsic to any CSS, namely the standard quantum limit $1/\sqrt{N}$. This leads to the condition $N \gtrsim q^2/(2\pi)^2$, valid for both the OAT as well as the dipolar dynamics.  

\section{Predictions of interferometry via parity measurements using tVMC}

The average parity $\langle P^z \rangle = \langle \prod_i (2S_i^z) \rangle$, as well as its variance, can be straightforwardly calculated via tVMC, giving that it is a diagonal quantity in the computational basis. In fact the dynamics is parity conserving, and given that $\langle P^z \rangle(0) = 0$ for the initial coherent-spin state, this remains valid at all times. Also $(P^z)^2 = 1$, so that ${\rm Var}(P^z) = 1$.  
As for the derivative of the average parity with respect to the rotation angle $\theta$ one has that
\begin{equation}
\frac{d\langle P^z \rangle}{d\theta} = - i \langle [P^z,J^x] \rangle =\langle  \sum_j (2S_j^y) \prod_{i\neq j} (2S_i^z) \rangle
\end{equation}
which can also be straightforwardly calculated via tVMC, for $\theta=0$.

\end{document}